\documentstyle[11pt,paspconf,epsf]{article}

\def\h{\noindent \hangindent=2.5pc }
\def\ts{\thinspace}
\def\um{$\mu$m}
\def\etal{{\it et al.}}
\def\arcs{$^{\prime\prime}$}
\begin{document}
\noindent{}

\noindent{\large \bf DEEP SUBMILLIMETER SURVEYS:}
\vspace{-10pt}

\noindent{\large \bf LUMINOUS INFRARED GALAXIES AT HIGH
REDSHIFT\footnote{Invited
	 review for the workshop {\it ``Space Infrared Telescopes and Related
	 Science"}, at the 32nd COSPAR Meeting, Nagoya, Japan, 1998, eds. T.
	 Matsumoto and T. de Graauw.  To be published in Advances in Space
	 Research (Oxford: Elsevier), 1999.}}

\vspace{0.3cm}

\noindent
{D. B. Sanders
\footnote{This paper draws on new material from 
the publication by Barger \etal\  1999, and the website
http:/www.ifa.hawaii.edu/$\sim$cowie.  Due to previous commitments,
Len Cowie and Amy Barger were unable to present this review.}}

\vspace{0.2cm}

\noindent{\it Institute for Astronomy, University of Hawaii,}\\
\noindent{\it 2680 Woodlawn Drive, Honolulu, HI\ 96822}\\
\noindent{\it Email: sanders@ifa.hawaii.edu}\\

\vspace{0.6cm}

\noindent{\bf ABSTRACT}

Deep surveys at 850\um\ from Mauna Kea using the SCUBA camera on the JCMT
appear to have discovered a substantial population of ultraluminous infrared
galaxies (ULIGs: $L_{\rm ir} > 10^{12} L_\odot$)\footnote{$L_{\rm ir} \equiv
L(8-1000)${\ts}$\mu$m in the object rest-frame.  Unless stated otherwise,
$H_{\rm o} = 75${\ts}km{\ts}s$^{-1}$Mpc$^{-1}$, $q_{\rm o} = 0${\ts}.} at high
redshift ($z \sim 1-4$). The cumulative space density of these sources
($\sim${\ts}$10^4${\ts} deg$^{-2}$ with $S_{850} > 1${\ts}mJy) is sufficient to
account for {\it nearly all} of the extragalactic background light at
submillimeter wavelengths.  Current estimates of the redshift distribution
suggest a peak in the comoving space density of SCUBA sources at $z = 1-3$,
similar to what is observed for QSOs and radio galaxies.  The luminosity
density in the far-infrared/submillimeter exceeds that in the UV by factors of
3--10 over this redshift range, implying that as much as 80--90{\ts}\% of the
``activity" in galaxies at $z \la 4$ is hidden by dust.  The SCUBA sources
plausibly represent the primary epoch in the formation of spheroids and massive
black holes triggered by major mergers of gas-rich disks.

\vspace{0.4cm}

\noindent{INTRODUCTION}

\vspace{0.2cm}

The Submillimeter Common User Bolometer Array (SCUBA) camera 
(Holland \etal\  1999) recently installed on the James Clerk Maxwell 
Telescope (JCMT) has provided a new window for ground-based 
studies of the high-$z$ Universe.  This brief review summarizes 
results from the large campaign of deep submillimeter 
surveys carried out during SCUBA's first year of operation on 
Mauna Kea ($\sim${\ts}Fall{\ts}'97--Summer{\ts}'98).  Evidence is presented 
that the SCUBA detections must be predominantly ULIGs at high redshift, and  
that the cumulative space density of these sources is sufficient to account 
for nearly 100{\ts}\% of the extragalactic background light (EBL) at 
far-infrared/submillimeter wavelengths.  Progress 
as well as current problems in identifying optical counterparts to the SCUBA 
sources are then discussed, followed by a review of model calculations 
comparing the luminosity density in the submm/far-IR with that in the UV 
over the redshift range, $z = 0-5$.  
Finally, we discuss evidence which suggests that the SCUBA sources, 
like local ULIGs, are powered by intense circumnuclear starbursts and 
powerful AGN both of which are fueled by major mergers of gas-rich disks.  

\vspace{0.2cm}

\noindent{SCUBA DEEP SURVEYS}

\vspace{0.2cm}

As this Workshop was being held, several observing teams were announcing
results from the first deep blank-field submillimeter surveys at
450\um/850\um\ using SCUBA.  Papers by Hughes \etal\  (1998) for the
UK-consortium and by Barger \etal\  (1998) for the Hawaii-Japan group were
announced in press releases during the week of the workshop, and results from
the Canada-France consortium were about to be submitted (Eales \etal\  1998).
These ``ultradeep" and ``deep" blank-field surveys follow the pioneering
surveys of Smail \etal\  (1997) who were the first to infer a substantial
population of luminous submillimeter galaxies from their SCUBA detections at
850{\ts}$\mu$m of background sources amplified by weak lensing from foreground
clusters.  With the exception of one source detected at 450{\ts}$\mu$m in the
cluster fields, all of the SCUBA deep field detections have been at 850$\mu$m.
The HDF (Hughes \etal\  1998), and Lockman Hole + SSA13 fields (Barger \etal\
1998a,b) are currently the three fields with the deepest SCUBA observations,
each with $\sim${\ts}50{\ts}Hrs total integration time reaching a noise level,
$\sigma_{850} \sim${\ts}0.6--0.8{\ts}mJy. Our results for the Lockman Hole are
discussed below.

\vspace{0.2cm}

\underbar{The Lockman Hole}

\vspace{0.2cm}

The 850{\ts}$\mu$m data on the Lockman Hole field (LH\_NW: 
Barger \etal\  1998, 1999) are shown in Figure 1.  The two SCUBA 
sources, LH\_NW1 and LH\_NW2, detected at 850{\ts}$\mu$m 
($>${\ts}3{\ts}$\sigma$), have 850{\ts}$\mu$m fluxes of 5.1{\ts}mJy, 
and 2.7{\ts}mJy, respectively, with upper limits at 450{\ts}$\mu$m 
of $\la${\ts}50{\ts}mJy (5{\ts}$\sigma$).

\begin{figure}[hbp] 
\plotone{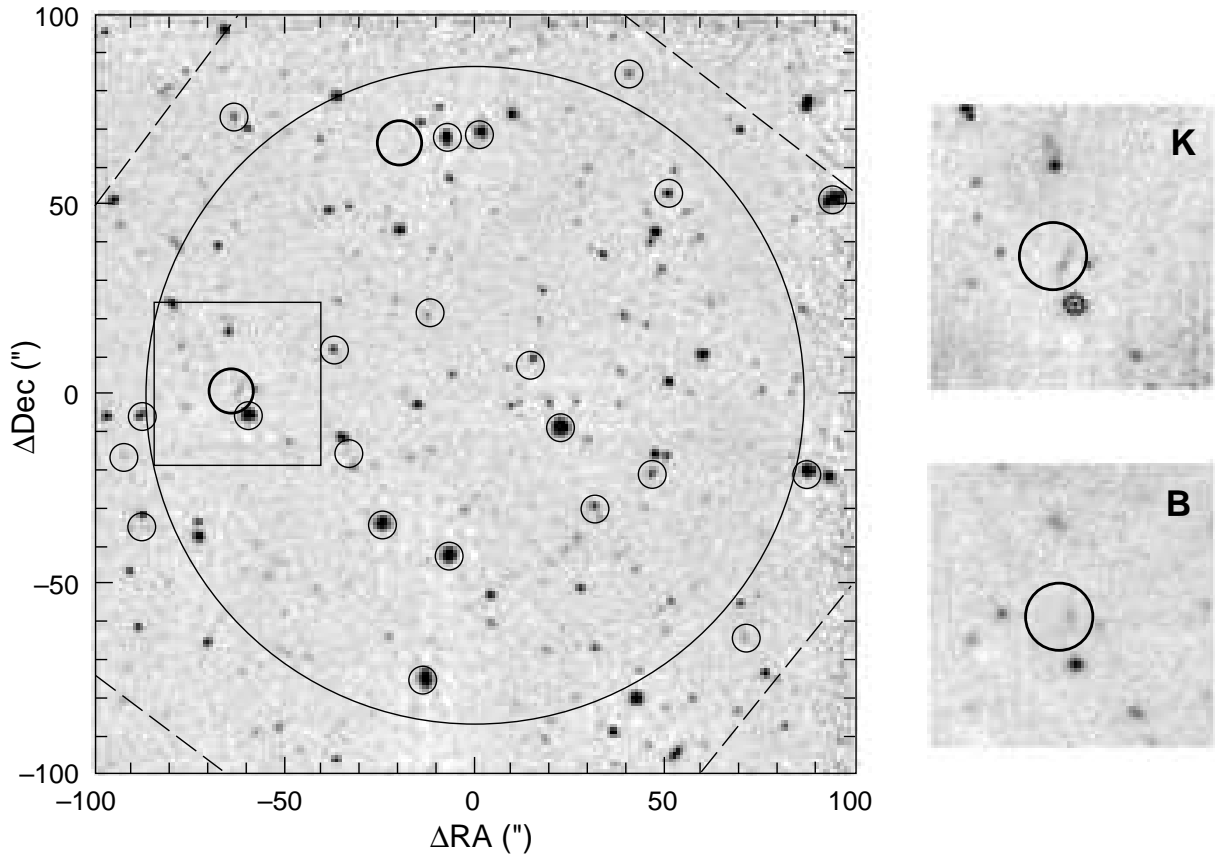}
\caption{SCUBA 850{\ts}$\mu$m detections (small thick circles: Barger \etal\
1998, 1999), and ISOCAM 7$\mu$m detections (small thin circles: Taniguchi
\etal\  1997) in the Lockman Hole northwest (LH\_NW) Deep Field (J2000: $RA =
10^h33^m55.5^s, Dec = +57^\circ46^\prime18^{\prime\prime}$) superimposed on a
K$^\prime$ image obtained with the QUick Infrared Camera (QUIRC) on the
University of Hawaii 88-inch telescope.  The field-of-view of the ISOCAM
detector array and the SCUBA array are indicated by a long dashed line and
large solid circle respectively.  On the right are two zoomed images of the
region outlined by the $45^{\prime\prime} \times 45^{\prime\prime}$ box which
is centered on the strongest SCUBA source   The zoomed K$^\prime$ image was
obtained with the Near InfraRed Camera (NIRC) on the Keck 10-m telescope, and
the zoomed B-band image was obtained with the University of Hawaii 2.2-m
telescope.} \end{figure}

Deep observations of the LH\_NW field have also been obtained 
in the mid- and far-infrared with {\it ISO}, at B and K$^\prime$ 
from Mauna Kea, and at 20{\ts}cm with the VLA.  As shown in Figure 1, 
neither SCUBA source has an ISOCAM 7{\ts}$\mu$m counterpart 
($< 35${\ts}$\mu$Jy; 5{\ts}$\sigma$).  LH\_NW1 appears to be 
centered on a faint K$^\prime$ source ($K_{\rm AB}^\prime = 21.8$) 
with disturbed morphology, which is barely detected in the current 
B-band image ($B_{\rm AB} = 23.5$).  LH\_NW2 is ``blank" in our K$^\prime$  
and B images implying that any counterpart has $K_{\rm AB}^\prime > 22.5$ 
and $B_{\rm AB} > 24.5$.  The ISOPHOT sources detected in a wider 
area ($44^\prime \times 44^\prime$) survey centered on LH\_NW have 
relatively large position uncertainties (typically 20\arcs--30\arcs), 
however there are no 95{\ts}$\mu$m or 175{\ts}$\mu$m detections 
within $\sim${\ts}2$^\prime$ of either SCUBA source (the 5{\ts}$\sigma$   
upper limits are $S_{95} < 40${\ts}mJy and $S_{175}< 75${\ts}mJy: 
Kawara \etal\  1998).  The VLA results are still preliminary, but any 
radio continuum counterparts must have $S_{\rm 20cm} \la 500${\ts}$\mu$Jy 
(5{\ts}$\sigma$) (M. Yun, private communication). 

\vspace{0.2cm}

\underbar{ULIGs at High Redshift}

\vspace{0.2cm}

From the strength of the 850{\ts}$\mu$m detections and the faintness of the
K$^\prime$ counterparts alone, it is relatively straightforward to show that
the SCUBA sources are most likely to be ULIGs at high redshift (i.e. $z
>${\ts}1).  The ``submillimeter excess", ($\equiv \nu S_\nu (850) / \nu S_\nu
(2.2)$), for both LH\_NW1 and LH\_NW2 is larger than 1 (2.4 and $>${\ts}3
respectively), which is impossible to produce from normal optically selected
galaxies at any redshift, or even by the most extreme infrared selected
galaxies at low redshift, but is almost exactly what would be expected for an
ULIG at high redshift.  Figure 2 shows that the expected flux for the
prototypical ULIG Arp{\ts}220 at $z >${\ts}1 is on the order of a few mJy at
850{\ts}$\mu$m. Also, the combination of a large negative K-correction in the
submillimeter plus a relatively flat or positive K-correction in the near-IR
naturally leads to values $\nu S_\nu (850) / \nu S_\nu (2.2) > 1$ for all ULIGs
at $z \ga${\ts}1.5{\ts}.  The observed faintness of the high-$z$ submillimeter
sources in current B-band images and the non-detections at 7{\ts}$\mu$m in the
deep ISOCAM images are consistent with the large U--B colors and the pronounced
minimum at $\sim${\ts}3-6{\ts}$\mu$m respectively, in the rest-frame SEDs of
ULIGs like Arp{\ts}220.  The non-detections at 95{\ts}$\mu$m and 175{\ts}$\mu$m
are consistent with the relatively high noise levels in the ISOPHOT images
coupled with the steep positive K-correction encountered over the rest-frame
wavelength range $\sim${\ts}20--80{\ts}$\mu$m for all ULIGs.

\begin{figure}[htbp]
\plotfiddle{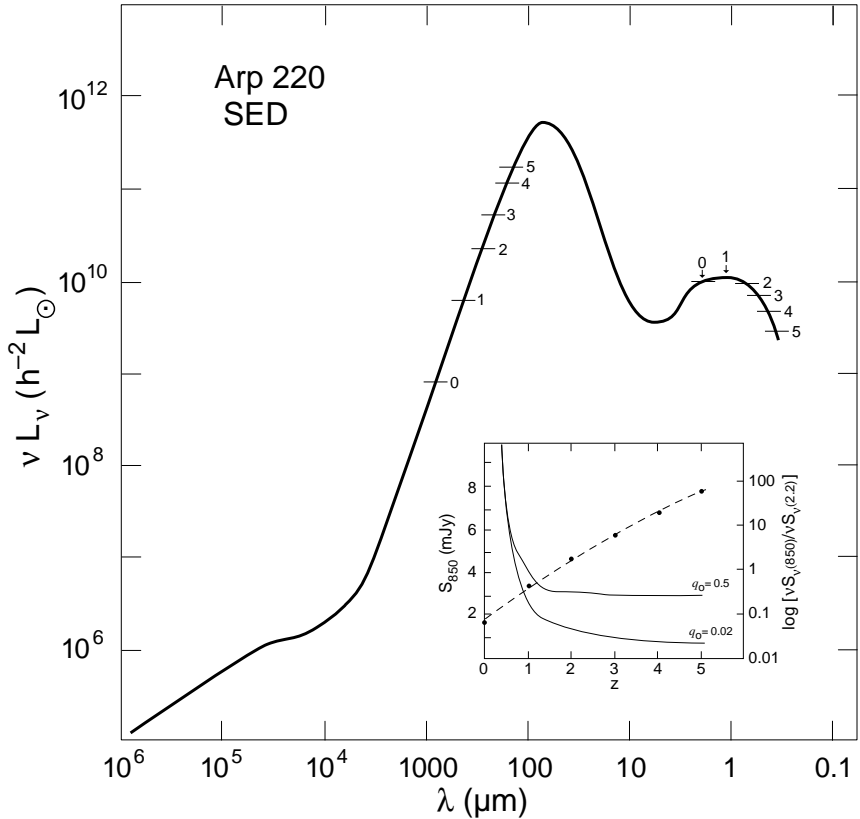}{4.3in}{0}{75}{75}{-230}{-120}
\caption{Observed radio-to-UV spectral energy distribution of the nearest ULIG,
Arp{\ts}220 ($z = 0.018$).  Labeled tickmarks represent object rest-frame
emission that will be shifted into the 850{\ts}$\mu$m and 2.2{\ts}$\mu$m
observed frame for redshifts, $z = 0 - 5$.  The insert shows the corresponding
observed-frame 850{\ts}$\mu$m flux and $\nu S_\nu(850)/\nu S_\nu(2.2)$ ratio
for Arp{\ts}220 at redshifts, $z = 0 - 5$.  } 
\end{figure}

\clearpage

\noindent{SOURCE COUNTS AT 850\um}

\vspace{0.2cm}

Table 1 summarizes the 850{\ts}$\mu$m cumulative source counts from all of the
SCUBA deep field surveys published to date. Taking into account the different
sensitivity limits and the reported formal uncertainties, there appears to be
good agreement on both the shape of the log$N$-log$S$ plot over the observed
range of 2--10{\ts}mJy, as well as on the extrapolated cumulative counts down
to 1{\ts}mJy obtained from $P(D)$ analyses of the low-level fluctuations in the
ultradeep images.  In retrospect, it is remarkable how little time it has taken
to reach consensus on the surface density of faint submillimeter sources per
SCUBA field.  This success is first and foremost due to the excellent
performance of SCUBA, but was also helped by a phenomenally well timed six
consecutive months of near record low mean submillimeter opacity on Mauna Kea,
and the commitment of relatively large blocks of observing time to several
teams of observers.

\begin{table*} \begin{center} \caption{SCUBA Source Counts at 850 microns$^a$}
\smallskip \begin{tabular}{cccll} \hline\noalign{\smallskip}
\hline\noalign{\smallskip} $\lambda$ & Counts & Flux Limit &Reference &Fields
\\ ($\mu$m) & (deg$^{-2}$) & (mJy) & & \\ \hline\noalign{\smallskip} 850 &
2500$\pm$1400 & 4 & Smail \etal\  (1997) & 2 clusters (A370, CL2244-02)\\
 "  &1100$\pm$600 & 8 & Holland \etal\  (1998) & background sources in 3 star
 fields\\ "  & 800$^{+1100}_{-500}$ & 3 & Barger \etal\  (1998)&LH\_NW; SSA13
 \\ "  & 7000$\pm$3000 & 1 & Hughes \etal\  (1998) & HDF \\ "  & 1800$\pm$600
 & 2.8 & Eales \etal\  (1998) & CFRS \\ "  & 1500$\pm$700  &4 & Blain \etal\
 (1999) & 7 clusters \\ "  & 7900$\pm$3000 & 1 &\qquad \qquad  "    & \qquad
 "   \\ "  & 10,000$\pm$2000 & 1 & Barger \etal\  (1999) & LH\_NW; SSA13;
 SSA17; SSA22\\ \hline\noalign{\smallskip} \hline\noalign{\smallskip}
\end{tabular} \end{center} \vspace{-0.3cm} \qquad \qquad $^a$ modified and
updated from a compilation by Trentham \etal\  (1998) \end{table*}

\begin{figure}[hbp]
\plotfiddle{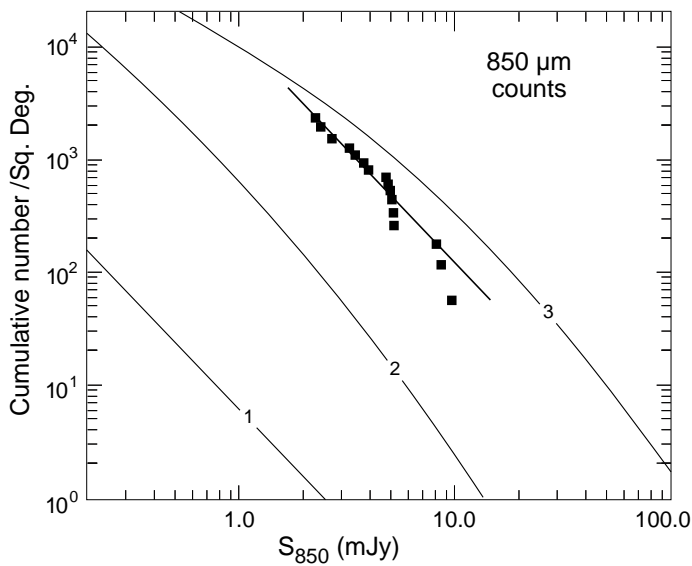}{3.2in}{0}{60}{60}{-190}{-120}
\caption{Comparison of the 850{\ts}$\mu$m source counts 
(solid squares: from Barger \etal\  1999) with semi-analytic model 
counts (see text).  Model 1 is equivalent to the ``no evolution" model 
of Blain \& Longair (1996).  Models 2 and 3 are equivalent to Models A and E 
respectively from Guiderdoni \etal\  (1998).}
\end{figure}

Figure 3 shows that the cumulative 850{\ts}$\mu$m 
counts from  2{\ts}mJy to 10{\ts}mJy can be approximated by a single 
power law of the form $N(>S) = 1 \times 10^4\ S^{-2}$ deg$^{-2}$. 
The extrapolated value at 1{\ts}mJy is consistent with the $P(D)$ 
results reported in Table 1.  This large surface density implies strong 
evolution in the ULIG population.    

Figure 3 compares the observed SCUBA counts with predictions from semi-analytic
models using three rather extreme distributions of ULIGs.  Model 1 is based on
the local IRAS 60{\ts}$\mu$m luminosity function of galaxies (i.e.
$\sim${\ts}0.001 ULIGs deg$^{-2}$ at $z <${\ts}0.08:  Kim \& Sanders 1998; see
also Soifer \etal\  1987; Saunders \etal\  1990) {\it assuming no evolution,
which underestimates the observed sources by nearly 3 orders of magnitude}.
Model 2 includes no ULIGs, but instead attempts to account for the fraction of
the optical/UV emission absorbed and reradiated by dust in sources observed in
optical/near-UV deep fields.  Model 2 still substantially underpredicts the
850{\ts}$\mu$m source counts by a factor of $\sim${\ts}30.  Model 3 includes a
strongly evolving population of ULIGs, constrained only by recent measurements
of the submillimeter background.  Model 3 is a much better match to the
observed SCUBA counts, although the particular redshift distribution assumed
(Model E of Guiderdoni \etal\  1998) was arbitrarily chosen to push most of the
ULIGs to $z >${\ts}3, which as described below is at odds with current redshift
estimates.

\vspace{0.2cm}

\noindent{CONTRIBUTION OF SCUBA SOURCES TO THE SUBMM/FAR-IR BACKGROUND}

\vspace{0.2cm}

Figure 4 compares the contribution of the 850{\ts}$\mu$m sources with the
recent model of the EBL determined from COBE data at far-infrared/submillimeter
wavelengths (Fixsen \etal\  1998; see also Hauser \etal\  1998; Puget \etal\
1996, 1998).  Approximately 25{\ts}\% of the 850{\ts}$\mu$m background resides
in sources brighter than 2{\ts}mJy, and {\it nearly all of the EBL at
850{\ts}$\mu$m can be accounted for by sources brighter than 1{\ts}mJy},
assuming the extrapolation down to 1{\ts}mJy given by the straight line fit to
the SCUBA data in Figure 3.

\begin{figure}[hbp]
\plotfiddle{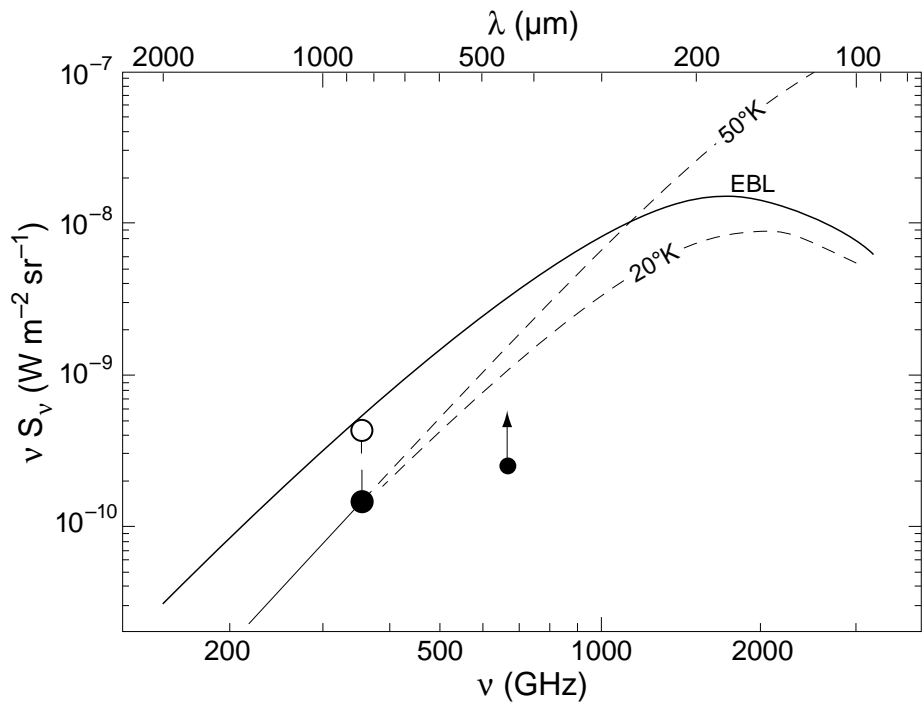}{3.8in}{0}{75}{75}{-230}{-170}
\caption{Comparison of the contribution of the 850{\ts}$\mu$m sources 
brighter than 3{\ts}mJy (solid circle) and extrapolated contribution 
of sources brighter than 1{\ts}mJy (open circle) to the EBL compared 
with the Fixsen \etal\ (1998) analytic approximation (solid curve) to the EBL. 
The two dashed curves are for observed source temperatures 
of 50{\ts}K and 25{\ts}K where each is based on a $\lambda$-weighted 
Planck function. (see http://www.ifa.hawaii.edu/$\sim$cowie/)}
\end{figure}

Figure 4 makes the important additional point that the average 
{\it observed} temperature of the 850{\ts}$\mu$m sources must 
be low (i.e. $\la${\ts}25{\ts}K) so that their emission will not 
exceed the EBL at wavelengths shortward of 200{\ts}$\mu$m.  If the 
SCUBA sources are indeed distant analogs of local ULIGs whose typical 
dust temperatures are $\sim${\ts}50--80{\ts}K, then a low observed 
temperature would provide additional evidence that the SCUBA sources 
lie predominantly at high redshift (i.e. $z >${\ts}1). 

\clearpage

\noindent{SOURCE IDENTIFICATION}

\vspace{0.2cm}

Progress in identifying optical/near-IR counterparts of the SCUBA deep-field
sources has been frustratingly slow, due in large part to the uncertainty in
the 850{\ts}$\mu$m SCUBA positions, but also to the intrinsic faintness of many
of the sources, even at K-band.  A prime example is the current controversy
surrounding the identified SCUBA counterparts in the HDF.  Four 4 of the 5
850{\ts}$\mu$m SCUBA sources in the HDF, the identifications by Hughes \etal\
(1998) hinge in large part on plausibility arguments using ``template SEDs".
Although these attempts are admirable, they apparently can be misleading, for
example  the identification of the brightest source, HDF850.1, with a faint
counterpart at $z = 3.4$ is now known to be in error from a more accurate
position obtained from millimeter interferometry.  Our results for the Lockman
Hole are not much more encouraging (see Figure 1):  LH\_NW1 seems to have an
identified K-band counterpart, however no emission lines are seen in a deep
Keck spectrum, and LH\_NW2 is a blank field in the current K-band image.

Progress has been somewhat better in the lensed cluster fields, where the 1-2
mags of amplification appears to reduce the number of blank fields ($\la
20${\ts}\% at $I < 25$: Smail \etal\  1998), but uncertainty in the
850{\ts}$\mu$m positions is still a major problem, plus contamination by
cluster members, unless properly accounted for, will artificially inflate the
number counts and skew the distribution of redshifts.  The wider-area shallower
surveys (e.g. Eales \etal\  1998; Lilly \etal\  1999) preferentially select
brighter nearby sources, which are more likely to have brighter optical/near-IR
counterparts, but still suffer from position uncertainty at 850{\ts}$\mu$m.

The net result after combining all current studies is that only
$\sim${\ts}20--25\% of the 850{\ts}$\mu$m sources currently have ``secure"
identifications, and even for these it has been necessary to rely heavily on
data at other wavelengths to make a strong case that only the identified source
has a high probability of being correct.

\vspace{0.2cm}

\noindent{TEMPLATE SEDs}

\vspace{0.2cm}

Attempts to sort out which of typically 2-3 equally probable sources is more
likely to be the true SCUBA counterpart have made use of ``template SEDs" to
determine which source is most likely to match the known properties of ULIGs.
This procedure works best when additional deep-field data are available at
several wavelengths, usually deep radio continuum images from the VLA, and high
resolution, deep near-IR and optical images from either {\it HST} or with
adaptive optics from the ground.  The most complete and deepest data sets are
of the HDF, with shallower multiwavelength data generally available for other
SCUBA fields.  However, even for the HDF, measurement uncertainties coupled
with a real dispersion in ULIG properties can still give ambiguous results.

Figures 5 and 6 illustrate the range of 
rest-frame SEDs observed for a sample of low-redshift ULIGs, and the 
corresponding flux ratios that would be observed for these objects 
at high redshift.  It is instructive to briefly review what has 
(or hasn't) been learned from trying to identify the correct SCUBA counterpart 
and/or to set meaningful constraints on redshift using observed flux ratios.

\begin{figure}[htbp]
\plotfiddle{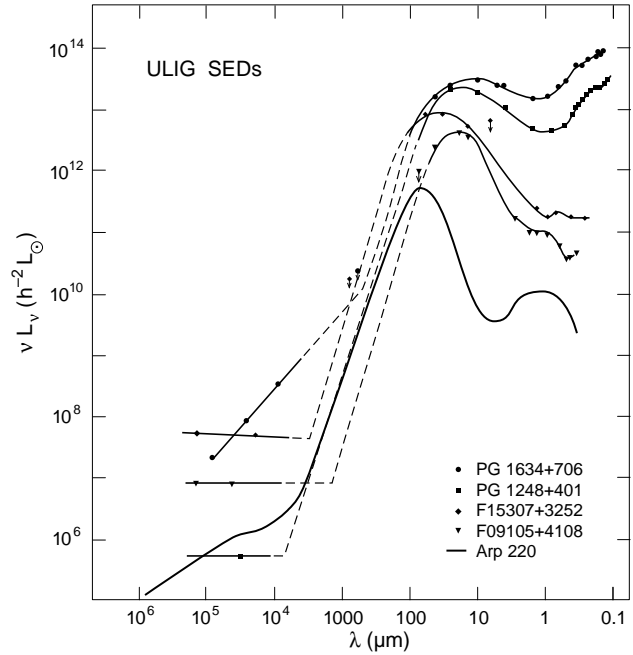}{3.5in}{0}{50}{50}{-150}{-80}
\caption{Measured rest-frame spectral energy distributions (SEDs) from 
UV-to-radio wavelengths for a representative sample of the most luminous 
sources detected by {\it IRAS}.}
\end{figure}

\begin{figure}[htbp]
\plotfiddle{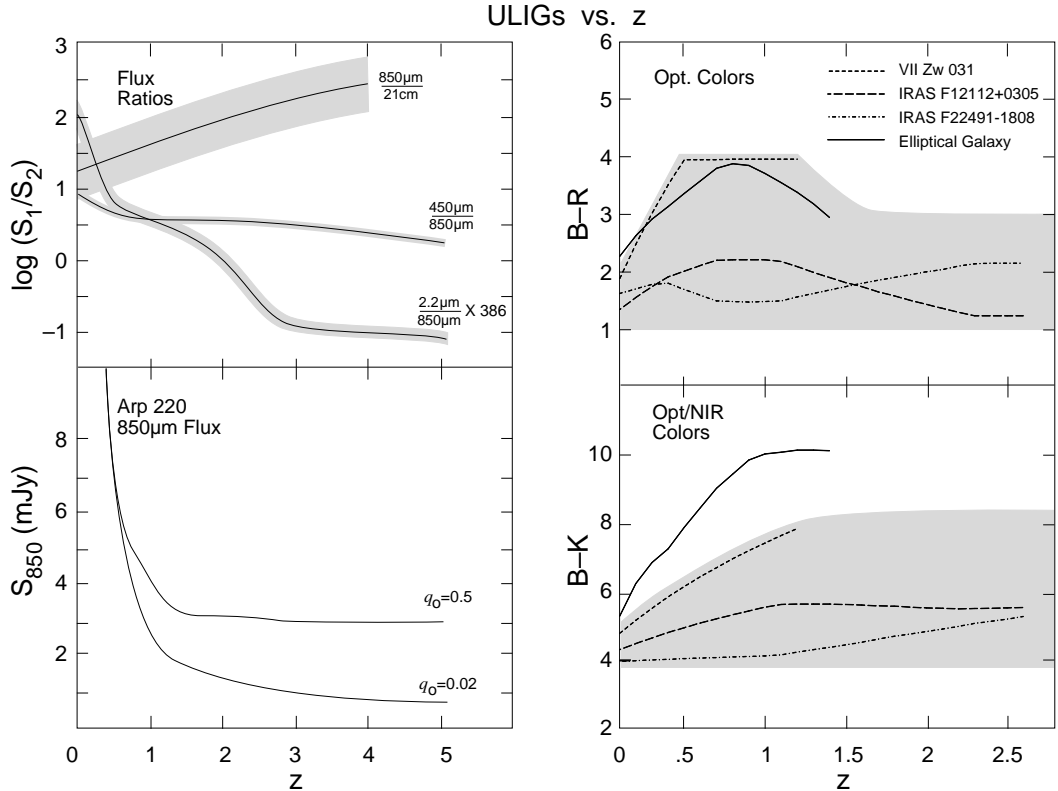}{4in}{0}{75}{75}{-230}{-120}
\caption{Left panels: (top) flux ratios as a function of redshift for 
nearby ULIGs from the {\it IRAS} Bright Galaxy Sample (Soifer \etal\  1987; 
Sanders \etal\  1995);\ (bottom) Flux versus redshift for Arp{\ts}220.
Right panels: (top) B--R versus redshift\ (bottom) B-K versus redshift \ 
for three nearby ULIGs (dashed/dotted lines) from the {\it IRAS} Bright 
Galaxy Sample compared with the redshifted colors of a standard elliptical 
galaxy (Trentham \etal\  1999).}
\end{figure}

\vspace{0.2cm}

\underbar{Radio /Far-Infrared Correlation}

Perhaps the most promising means of identifying SCUBA positions is from high
resolution radio continuum maps, as has been done quite successfully for lower
redshift {\it IRAS} sources (e.g. Condon \etal\  1990).  The subarcsecond radio
positions coupled with a relatively tight correlation between the radio and the
far-infrared emission for infrared galaxies (e.g. Condon \etal\  1991) could
provide relatively unambiguous identifications of the current SCUBA deep field
sources, provided that it is possible to reach a point source sensitivity of
$\la 10${\ts}$\mu$Jy at 3cm or $\la 30${\ts}$\mu$Jy at 20cm.  Unfortunately,
although this has been achieved for the HDF (Richards \etal\  1998), {\it the
majority of the SCUBA sources in the HDF are not detected in the radio
continuum at a level which is already at least 3--5 times fainter than even the
maximum observed ratios shown on the plot in Figure 6}.  One possible
explanation would be to push the majority of the SCUBA sources to very large
redshifts, i.e. $z > 3$ (e.g. Carilli and Yun 1999), but as shown below, this
is in conflict with other best estimates of the range of redshift for the
majority of the SCUBA sources. The current situation is bizarre enough to have
caused others to attempt interesting explanations such as a substantial
underestimate of the pointing error in the JCMT observations of the HDF
(Richards 1999).  A more likely explanation is that the SCUBA sources are
indeed fainter in the radio than ULIGs studied by {\it IRAS}, although why this
should be is currently not clear.  The bottom line is that ultradeep VLA
observations are currently incapable of detecting the majority of SCUBA deep
field sources, and thus even deeper ultradeep VLA observations are needed if
there is to be any hope of using this potentially powerful diagnostic to pin
down the SCUBA positions.

\underbar{450\um/850\um\ Spectral Index}

In all of the SCUBA deep-fields, only one source has been reliably detected
(5$\sigma$) at 450{\ts}$\mu$m, that being SMM02399-0136 in the field of the $z
= 0.37$ cluster, A370; it is the brightest 850{\ts}$\mu$m source detected in
the survey of Smail \etal\  (1998), and by far, the brightest SCUBA source yet
detected in any of the blank field surveys.  For all other sources only upper
limits can be used to constrain the 450\um/850\um\ spectral index.  Given that
the noise at 450{\ts}$\mu$m is typically 10$\times$ that at 850{\ts}$\mu$m
current limits on this ratio are simply not sensitive enough to be of much
use.  The only caveat might be the HDF where the exceptionally good weather
produced a quoted limit at 450{\ts}$\mu$m of $\sim${\ts}25{\ts}mJy (Hughes
\etal\  1998), implying that the two brightest 850{\ts}$\mu$m sources have
$S_{450}/S_{850} < 6-7$; however, this sets only a weak constraint of $z \ga
0.3$ (see Figure 6).

\underbar{K-band versus 850\um\ Flux}

Figure 6 shows the effects of the large negative K-correction in the
submillimeter on the ``submillimeter excess" ratio ($\equiv \nu S_\nu (850)/\nu
S_\nu (2.2)$).  This ratio provides a useful redshift discriminant at $z \la
3$, however at higher redshift the fact that the large positive K-corrections
in both the mid-infrared and in the near-UV are similar means that there is
little variation in the ratio over the redshift range, $z \sim 3-6$.

\underbar{Optical, Near-Infrared Colors}

One of the surprises from the early results from the lensed cluster survey was
the finding by Smail \etal\  (1998) that the optical ($V-I$) colors of the
SCUBA counterparts (using only those sources with the highest probability of
being correctly identified) are {\it bluer} on average than the faint field
population, although the spread in colors is relatively large and at least a
few objects appear to be extremely red.  A priori one might have expected that
the majority of the SCUBA sources, given their large infrared excess, should be
extremely red objects (e.g. EROs: Hu and Ridgeway 1991; Cimatti \etal\  1998;
Dey \etal\  1998).  Although ULIGs are faint at optical wavelengths, either
enough UV light escapes from the nuclear dust shroud or is still produced by a
slowly decaying starburst in the surrounding disk.  Only in the most extreme
geometries is it apparently possible to hide all evidence of current and past
starburst activity. Unfortunately, given that the optical colors of SCUBA
sources bracket those of the faint field population, optical colors may only be
of marginal help in identifying SCUBA counterparts.

\underbar{$L_{\rm fir} / L_{\rm CO}$}

This ratio differs from the other template measures in that it relies on a line
measurement, and thus is only useful once a redshift has been determined.  The
great majority of ULIGs (at all redshifts) appear to have molecular gas masses
in the range $M({\rm H}_2) = 0.5-10 \times 10^{10} M_\odot$ as inferred from
their luminosities in one or more of the lower rotational transitions of CO
(e.g. Sanders \etal\  1991; Downes \etal\  1995; Solomon \etal\  1997).  The
sensitivity of current millimeterwave interferometers is sufficient to detect
$\ga 10^{10} M_\odot$ of molecular gas independent of redshift (see Radford
1994), and thus once a redshift has been determined, detection of the source in
CO would greatly increase the probability that it indeed was the source of the
submillimeter emission.  The only drawback is that ULIGs with ``warmer" than
normal mid- and far-infrared colors can have substantially less molecular gas,
e.g. $M({\rm H}_2) < 10^9 M_\odot$ (Evans \etal\  1998), and thus fall well
below current sensitivity limits.  Therefore, given that the mid-infrared
colors of SCUBA sources are not well-known, non-detection in CO would not
necessarily rule out the identification of an otherwise reasonable
counterpart.

Despite the above attempts to identify the 850{\ts}$\mu$m deep-field SCUBA
sources, most still do not have secure identifications.  Uncertainty in the
850{\ts}$\mu$m positions remains the major problem.  Until submillimeter
interferometers start to come on line early in the next decade, the most useful
data to obtain would be interferometer measurements at 1{\ts}mm to detect the
long wavelength tail of the 850{\ts}$\mu$m emission.  Unfortunately the
expected 1{\ts}mm flux for most of the deep-field SCUBA sources is $\la
1${\ts}mJy, which is below the practical sensitivity limits of current mm-wave
interferometers. Deeper ``ultra-deep" VLA surveys may continue to be the most
promising alternative.

\clearpage

\noindent{REDSHIFT DISTRIBUTION}


Clearly the most important unknown concerning the SCUBA sources is their
redshifts.  The relatively small fraction ($\sim${\ts}20--25{\ts}\%) of SCUBA
sources that have secure identifications has resulted in much speculation about
the true redshift distribution.  Although it is not possible to rule out
distributions that peak at $z \ga 3$ as originally advocated by Hughes \etal\
(1998) for the HDF, or that peak at $z \sim 1$ as advocated by Eales \etal\
(1999) and Lilly \etal\  (1998) for the CFRS fields, there is increasing
evidence that the comoving luminosity function of SCUBA sources peaks somewhere
in between, i.e. at $z \sim 1-3$.

The most comprehensive redshift survey of SCUBA sources is that for the 16
sources in the lensed cluster fields of Smail \etal\  (1998), as recently
reported in Barger \etal\  (1999).  Of the 16 sources, two are clearly cluster
contaminants (i.e. cD galaxies).  Of the remaining 14 those with the strongest
identifications are an interacting pair at $z = 2.55$ and the previously
identified (Ivison \etal\  1998) type-II AGN/starburst source at $z = 2.8$. The
next two most reliable detections, based partly on template SEDs, are at $z =
1.06$ and $z = 1.16$.  Eight sources are still highly uncertain, however nearly
all plausible nearby counterparts have redshifts in the range $z = 0.18 -
2.11$, with two galaxies at the low end of this range likely being cluster
contaminants.  Finally, only two sources have no obvious optical counterparts,
and therefore might be at higher redshifts (i.e. $z > 3$).


\noindent{THE ``STAR FORMATION HISTORY" OF THE UNIVERSE}


The past few years has seen a revolution in optical studies 
of field galaxies, which with the help of 8-10{\ts}m-class optical telescopes,  
has enabled the determination of the UV luminosity density of the 
Universe out to redshifts $z \ga 4$.  

\begin{figure}[hbp]
\plotfiddle{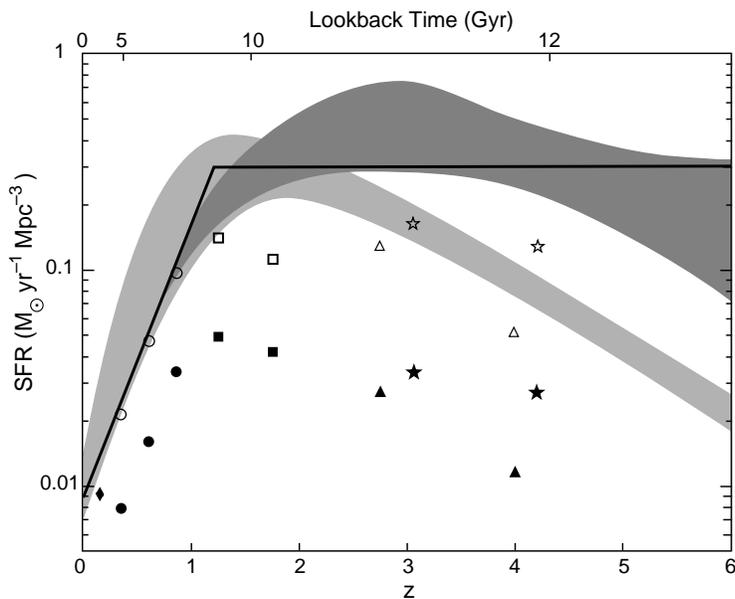}{2.5in}{0}{55}{55}{-175}{-120}
\caption{The global ``star formation history" 
of the Universe ($H_{\rm o} = 50${\ts}km{\ts}s$^{-1}$Mpc$^{-1}$, 
$q_{\rm o} = 0.5$ is used for consistency with Madau \etal\  1996).  
In the optical/near-UV, the mean co-moving rate of star 
formation is determined from the total measured rest-frame  
UV luminosity density of galaxies (solid hexagon: Trayer \etal\  1998; 
solid circles: Lilly \etal\  1996; 
solid squares: Connolly \etal\  1997; solid triangles: Madau \etal\  
1996; solid stars: Steidel \etal\  1998).  The optical data points have 
been corrected upward for extinction; 
the open symbols shown here are the correction ($\times 4.7$) 
adopted by Steidel \etal\  
(1998).  The light shaded region represents a prediction from 
chemical evolution models (Pei and Fall 1995).  The dark shaded 
region and thick solid line represent the maximum contribution from 
far-IR/submm sources (i.e. assuming all of the far-IR/submm 
emission is powered by young stars) using models with a range of 
$z$ distributions which are consistent with the current 
observations of 850{\ts}$\mu$m SCUBA sources (Blain \etal\  1998).}
\end{figure}

Figure 7 gives the optical/near-UV determination of the mean co-moving rate 
of star formation over the redshift range $z = 1-4$ following an early 
analysis of galaxies in the HDF by Madau \etal\  (1996),  
who also noted that there seemed to be little evidence for substantial 
amounts of dust obscuration in the optical sources. It is remarkable how 
quickly this picture has changed.

It is immediately obvious from the large space density of the 850{\ts}$\mu$m
sources, and their large inferred bolometric luminosity in the
far-IR/submillimeter (i.e. corresponding to $\sim 30-300 L_{\rm bol}^\ast$),
that the SCUBA sources could represent a significant and perhaps dominant
component of the luminosity emitted by galaxies over cosmic time.  It is also
clear that a substantial fraction of the ``activity" in galaxies at high
redshifts ($z >${\ts}1) is obscured by dust, and, therefore has been missed in
optical/UV surveys.  This is illustrated in Figure 7, which assuming that all
of the far-IR/submm luminosity is due to star formation, plots a range of
plausible models of SFR vs. $z$ (Blain \etal\  1998) consistent with reasonable
assumptions for the redshift distribution of the SCUBA sources.  Figure 7
suggests that the SCUBA sources dominate the optical by factors of 5--10 at
nearly all redshifts, and that their distribution with redshift is likely to be
substantially different than the bias toward low redshifts inferred from the
early optical studies.

However, it is naive to assume that the role of dust is negligible in the
optical sources, and indeed from studies of the optical/near-UV colors it was
quickly realized that dust obscuration could have a substantial effect.
Reddening models developed for local starburst galaxies (e.g. Meurer \etal\
1997, 1998; Calzetti 1997, 1998) have now been used to correct the optical
sources for reddening, resulting in mean model-dependent increases of
$\times$3--7 in the SFR (e.g. Dickinson 1998; Pettini \etal\  1998; Steidel
\etal\  1998).  Better statistical samples at $z > 3$ have also erased evidence
of a turnover in the distribution at $z \sim 1$ (Steidel \etal\  1998), and
with an additional assumption that mean reddening corrections are relatively
constant (at least over the observable redshift range out to $z \sim 4$:  e.g.
Steidel \etal\  1998), the redshift distribution of star formation from the
optical samples begins to resemble more closely that determined for the SCUBA
sources, although the latter still exceed the former by factors of $\sim$2--4
at $z > 1$.

\vspace{0.2cm}

\noindent{THE STARBURST--AGN CONNECTION}

\vspace{0.2cm}

Are the SCUBA sources simply powered by starbursts as is the case for 
the optical-UV selected samples, or could AGN play a significant role in 
producing their enormous luminosities{\ts}? 

Although relatively little is currently known about the detailed properties of
the SCUBA sources, it is interesting to note that the bulk properties of those
few sources whose positions and redshifts have been reliably identified with
SCUBA detections are in fact quite similar to the mixture of starburst and AGN
properties, and merger morphologies observed for ULIGs at low redshift.

The two best studied sources from the sample of Smail \etal\  (1997) are
illustrative.  SMM{\ts}J02399$-$0136 at $z \sim 2.8$, with $L_{\rm ir} \ga
10^{13} L_\odot$, is morphologically compact with an optical classification as
a narrow-line ``type-2" AGN (Ivison \etal\  1998), and contains
$\sim${\ts}$10^{10.5} M_\odot$ of molecular gas (Frayer \etal\  1998).
SMM{\ts}J14011+0252 at $z \sim 2.6$ with $L_{\rm ir} \sim 10^{12.3} L_\odot$
(Barger \etal\  1999), is a strongly interacting/merger pair, with an
H{\ts}II-like optical spectrum and $\sim${\ts}$10^{10.7} M_\odot$ of molecular
gas (Frayer \etal\  1999).  These two sources fit remarkably well into the
pattern exhibited by ULIGs in the local Universe.  In particular their
molecular gas masses, optical luminosities, and optical morphologies are very
similar to what is observed for local ULIGs (see Sanders and Mirabel 1996 for a
review of local ULIGs).

Although there are only two objects, it is interesting to note that their
optical spectral types fit the trend with $L_{\rm ir}$ that is observed for
local ULIGs, as illustrated in Figure 8 which shows that the importance of AGN
appears to increase with increasing $L_{\rm ir}$.  At $L_{\rm ir} >
10^{12.0-12.3} L_\odot$ 10--20{\ts}\% of local ULIGs appear to be predominantly
powered by an AGN with the remainder apparently powered by a starburst
(although it is still not possible to rule out that a substantial fraction of
the luminosity is produced by a heavily dust enshrouded AGN).  At $L_{\rm ir} >
10^{12.3} L_\odot$ nearly half of all local ULIGs appear to be powered
predominantly by an AGN, while at the highest observed infrared luminosities
(i.e. $L_{\rm ir} \ga 10^{13} L_\odot$) all of the currently identified objects
are classified as AGN, typically as Sy{\ts}2s in direct optical emission (e.g.
IRAS{\ts}F09105+4108: Kleinmann and Keel 1987; IRAS{\ts}F15307+3252: Cutri
\etal\  1994; IRAS{\ts}F10214+4724: Rowan-Robinson \etal\  1991), but have been
shown to contain hidden broad-line regions in polarized optical light (e.g.
Hines \etal\  1995), or in direct near-infrared emission (e.g. Veilleux \etal\
1997, 1998).

\begin{figure}[htbp]
\plotfiddle{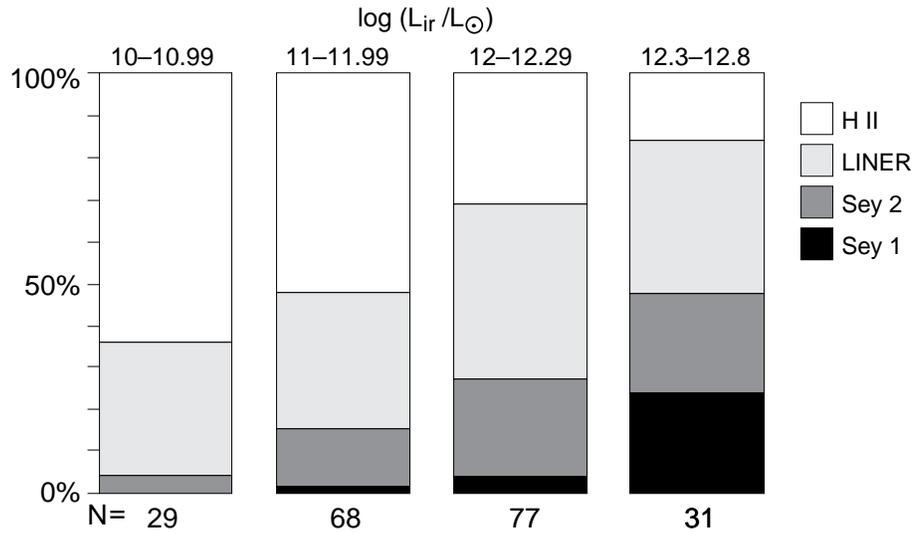}{2.5in}{0}{150}{150}{-460}{-540}
\caption{The optical spectral classification of infrared galaxies 
versus infrared luminosity (Veilleux \etal\  1999).}
\end{figure}

Although Arp{\ts}220 (optically classified as a LINER) is often used as the
template for ULIGs (see Figure 2) primarily due to the fact that it is the
nearest such object by nearly a factor of 2.5, it is not necessarily typical of
most UlIGs, in particular those with the largest infrared luminosities.
Mrk{\ts}231 (optically classified as a Sy{\ts}1/BALQSO) may better illustrate
the mixture of starburst and AGN found in many ULIGs.

Figure 9 shows the merger morphology apparent in deep optical images of
Mrk{\ts}231, and the detection by {\it HST} of a large population of star
clusters in the inner 1-3 kpc, presumably the remnants of a more widespread
circumnuclear starburst whose total bolometric luminosity is currently only a
very small fraction of the total bolometric luminosity of the system (Surace
\etal\  1998), nearly all of which originates from the inner few hundred
parsecs (e.g. Matthews \etal\  1987).  The dense knot of molecular gas in the
inner $\sim${\ts}1{\ts}kpc which is centered on the Sy{1} nucleus, presumably
represents the large reservoir of gas still available to fuel the last phases
of the circumnuclear starburst as well as to continue to build and fuel the
AGN.

\begin{figure}[htbp]
\plotfiddle{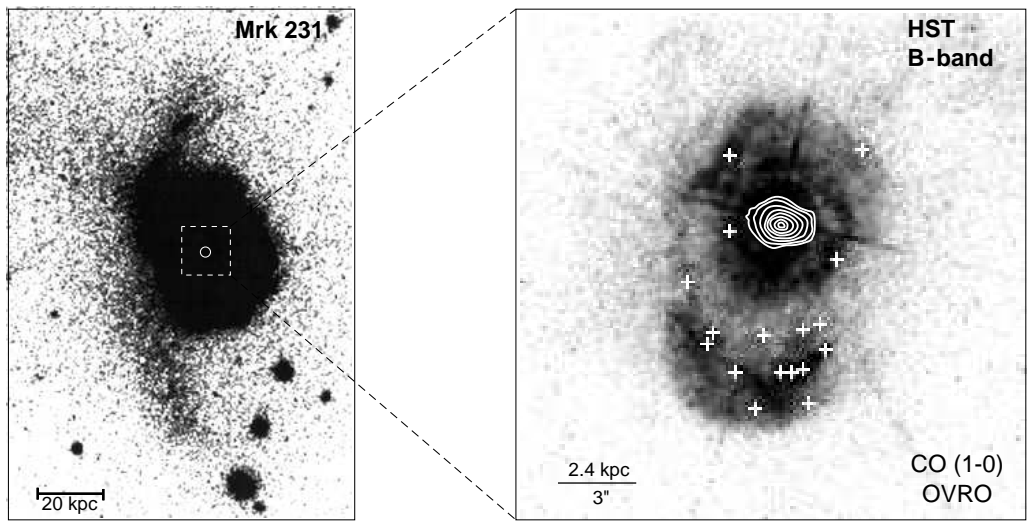}{3in}{0}{100}{100}{-200}{0}
\caption{The advanced merger/ULIG/QSO Mrk231 -- Left panel: optical 
image (Sanders \etal\  1987) and CO contour (Scoville \etal\  1989).  
Right panel: {\it HST} B-band image and identified stellar 
clusters (`+') from Surace \etal\  (1998).  The high resolution 
CO contours are from Bryant and Scoville (1997).}
\end{figure}

\clearpage

\noindent{FINAL COMMENTS}

\vspace{0.2cm}

One of the major questions at this Workshop has been the relationship of the
SCUBA sources to the optically selected high-$z$ population of starburst
galaxies.  One view is that the SCUBA sources are indeed just the most heavily
reddened objects already contained in the optical samples.  Favoring this view
is the evidence (summarized by Steidel \etal\  1998) that on average the more
luminous objects in optical samples are also redder, such that after correction
for extinction they would have intrinsic luminosities equivalent to that of the
SCUBA sources (i.e. $\ga 10^{12} L_\odot$).  
detect an 
emerged 
``star-formation histories" derived for the two populations {\it may have}
similar shape might suggest that both samples are simply different
manifestations of the same general phenomenon.

An alternative view is that the SCUBA sources represent an inherently distinct
population, for example the formation of massive spheroids and massive black
holes, both of which are triggered by the merger of two large gas-rich disk
galaxies (e.g. Kormendy and Sanders 1992; Kormendy and Richstone 1995; see also
the review in Sanders and Mirabel 1996).  Favoring this view is the fact that
the peak in the SCUBA redshift distribution at $z = 1-3$ is similar to what is
observed for QSOs (Schmidt \etal\  1995; Warren \etal\  1994; Shaver \etal\
1998) and radio galaxies (Dunlop 1997).  The relatively flat redshift
distribution now infered for the UV starburst population at $z > 1$ (Steidel
\etal\  1998) might then better represent the building of gas-rich disks over a
wider range of cosmic time.  Also, there is no current evidence to show that
the SCUBA detections are indeed related to the most heavily reddened optical
sources, plus it is not clear that the far-UV luminosity of SCUBA sources is
sufficient to be included in the deepest HST samples,
 i.e. $L_\nu (2200\AA) / L_\nu (80\mu m) \sim 10^{-5.5}$, for Arp{\ts}220-like
objects (see Figure 2).

Finally, in response to questions raised during the AGN Workshop at this COSPAR
meeting, we consider the intruiging possibility that the large population of
newly discovered SCUBA sources might be responsible for producing a large
fraction of the X-Ray background.  It is now aparent from the shape of the
X-Ray background spectrum that most of the objects contributing to the X-Ray
background are heavily absorbed (e.g. Fabian and Barcons 1992; Boyle \etal\
1995; Almaini \etal\  1998), and that they have largely been missed in optical
surveys.  Candidates for these obscured sources are the class of ``narrow-line
X-Ray galaxies" (NLXGs: Hassinger 1996), which have recently been characterized
by Maiolino \etal\  (1998) has having extremely heavy obscuration along the
line of sight (e.g. $N_{\rm H} > 10^{24-25} {\rm cm}^{-2}$).  SCUBA sources
(like local ULIGs) are obvious candidates for having such large absorbing
columns, and if a significant fraction of the SCUBA sources prove to have
buried AGN, then it seems reasonable to assume that they might indeed be
responsible for producing a substantial fraction of the X-Ray background.  Deep
surveys with soon-to-be-launched X-Ray satellites (e.g. {\it XMM}, {\it
Chandra}, {\it Astro-E}) could provide the answer.

\vspace{0.4cm}

\acknowledgments

I am grateful to Karen Teramura for assistance in preparing the 
figures, and to JPL contract no. 961566 for partial financial support.

\end{document}